\documentclass[fleqn,usenatbib]{mnras}

\usepackage{newtxtext,newtxmath}

\usepackage[T1]{fontenc}

\DeclareRobustCommand{\VAN}[3]{#2}
\let\VANthebibliography\thebibliography
\def\thebibliography{\DeclareRobustCommand{\VAN}[3]{##3}\VANthebibliography}

\usepackage{graphicx}	
\usepackage{amsmath}	

\usepackage{xspace}
\usepackage{url}
\usepackage{xcolor}
\usepackage{cjhebrew}
\usepackage{fontawesome}
\newfam\hebfam
\font\tmp=rcjhbltx at10pt \textfont\hebfam=\tmp
\font\tmp=rcjhbltx at7pt  \scriptfont\hebfam=\tmp
\font\tmp=rcjhbltx at5pt  \scriptscriptfont\hebfam=\tmp
\edef\declfam{\ifcase\hebfam 
	0\or1\or2\or3\or4\or5\or6\or7\or8\or9\or A\or B\or C\or D\or E\or F\fi}
\mathchardef\tav   = "0\declfam 74

\newcommand{\code}[1]{\texttt{#1}\xspace}
\newcommand{\org}[1]{\textsc{#1}\xspace}
\newcommand{\run}[1]{\textit{#1}\xspace}

\newcommand{\desc}{\org{DESC}}
\newcommand{\lsst}{\org{LSST}}
\newcommand{\pz}{photo-$z$\xspace}
\newcommand{\pzs}{photo-$z$s\xspace}

\newcommand{\pzpdfs}{photo-$z$ posteriors\xspace}
\newcommand{\os}{OS\xspace}
\newcommand{\oss}{OSs\xspace}

\newcommand{\opsim}{\code{OpSim}}
\newcommand{\maf}{\code{MAF}}
\newcommand{\healpix}{\code{HEALpix}}
\newcommand{\cmnn}{\code{CMNN}}
\newcommand{\pzflow}{\code{pzflow}}
\newcommand{\jax}{\code{Jax}}

\newcommand{\tlmcode}{\code{TheLastMetric}}
\newcommand{\tlmsymb}{\ensuremath{\tav}\xspace}
\newcommand{\tlmword}{TLM\xspace}
\newcommand{\data}{\ensuremath{x_{phot}}}
\newcommand{\E}{\mathbb{E}}

\newcommand{\base}{\run{baseline\_v1.5}}
\newcommand{\fpstuck}{\run{footprint\_stuck\_rolling\_v1.5}}
\newcommand{\ddf}{\run{ddf\_heavy\_nexp2\_v1.6}}

\newcommand{\bare}{\run{barebones\_v1.6}}

\newcommand{\nblink}[1]{\footnote{\url{https://github.com/aimalz/TheLastMetric/blob/master/#1.ipynb}}}
\newcommand{\github}{\footnote{\url{https://github.com/aimalz/TheLastMetric}}}

\title[An information-based metric of photometric redshifts for observing strategy optimization]{An information-based metric for observing strategy optimization, demonstrated in the context of photometric redshifts with applications in cosmology}

\author[Malz, Lanusse, Crenshaw \& Graham]{Alex I. Malz,$^{1}$\thanks{E-mail: aimalz@astro.ruhr-uni-bochum.de}
Francois Lanusse,$^{2}$
John Franklin Crenshaw,$^{3}$
\& Melissa L. Graham$^{4}$
\\
$^{1}$Ruhr-University Bochum, German Centre for Cosmological Lensing, Universit\"{a}tsstra{\ss}e 150, 44801 Bochum, Germany\\
$^{2}$AIM, CEA, CNRS, Universit\'e Paris-Saclay, Universit\'e Paris Diderot, Sorbonne Paris Cit\'e, F-91191 Gif-sur-Yvette, France\\
$^{3}$Department of Physics, University of Washington, Box 351560, Seattle, WA 98195\\
$^{4}$DiRAC Institute, Department of Astronomy, University of Washington, Box 351580, U.W., Seattle, WA 98195, USA\\
}

\date{Accepted XXX. Received YYY; in original form ZZZ}

\pubyear{2021}

\begin{document}
\label{firstpage}
\pagerange{\pageref{firstpage}--\pageref{lastpage}}
\maketitle

\begin{abstract}
  The observing strategy of a galaxy survey influences the degree to which its resulting data can be used to accomplish any science goal.
  LSST is thus seeking metrics of observing strategies for multiple science cases in order to optimally choose a cadence.
  Photometric redshifts are essential for many extragalactic science applications of LSST's data, including but not limited to cosmology, but there are few metrics available, and they are not straightforwardly integrated with metrics of other cadence-dependent quantities that may influence any given use case.
  We propose a metric for observing strategy optimization based on the potentially recoverable mutual information about redshift from a photometric sample under the constraints of a realistic observing strategy.
  We demonstrate a tractable estimation of a variational lower bound of this mutual information implemented in a public code using conditional normalizing flows.
  By comparing the recoverable redshift information across observing strategies, we can distinguish between those that preclude robust redshift constraints and those whose data will preserve more redshift information, to be generically utilized in a downstream analysis. 
  We recommend the use of this versatile metric to observing strategy optimization for redshift-dependent extragalactic use cases, including but not limited to cosmology, as well as any other science applications for which photometry may be modeled from true parameter values beyond redshift. 
  \github
\end{abstract}

\begin{keywords}
surveys -- galaxies: distances and redshifts -- methods: statistical
\end{keywords}



\section{Introduction}
\label{sec:intro}

The Vera C. Rubin Observatory will produce a catalog of tens of billions of astronomical objects over the course of the ten-year Legacy Survey of Space and Time (\lsst; \citealt{2019ApJ...873..111I}).
The quality and quantity of resulting data will depend on \lsst's observing strategy (\os), which encompasses the choice of frequency and duration of visits to each portion of the night sky across each of \lsst's $ugrizy$ filters as a function of the survey's duration.
As the \os directly impacts the science one can accomplish with the resulting data \citep{jones_survey_2020}, \lsst's Science Collaborations (SCs) are directing considerable effort to optimizing the choice of \os \citep[e.g.,][to name but a few]{2017arXiv170804058L, graham_photometric_2018, lochner_impact_2021}.

Though the space of all \oss is very high-dimensional, a decision of \os may be informed by how each science goal is affected by each \os considered.
\lsst has developed two tools\footnote{\url{https://pstn-051.lsst.io/}} to facilitate an optimal choice of \os.
\opsim \citep{lsst_opsim_2016, delgado_lsst_2014} forecasts the impact of an \os on the properties of the resulting photometric observations \lsst will deliver.
The \maf (Metrics Analysis Framework) \citep{lsst_maf_2017} was established to ensure that the choice of \os would be well-informed by all science cases, whose proponents are invited to include one or more metrics to be automatically evaluated for each set of simulated observational conditions.
Though each science application may favor a different \os, a fair choice may be made by reviewing how all science goals are affected.

The utility of a gargantuan catalog of extragalactic objects, such as what \lsst will provide, relies on the accuracy and precision of its constraints on their redshifts, which, for a photometric survey such as \lsst, represent a dominant factor in the error budgets of most if not all extragalactic science applications.
Without access to high-fidelity spectroscopic redshift measurements, users of \lsst's extragalactic catalog will rely on photometric redshift (\pz) estimates, which suffer from multiple forms of uncertainty, even under idealized conditions \citep[see][and extensive references therein]{schmidt_evaluation_2020}.

Redshift uncertainty thus represents one of \lsst's greatest liabilities and one of the utmost importance to multiple \lsst SCs \citep{awan_testing_2016, lochner_optimizing_2018, scolnic_optimizing_2018, almoubayyed_optimizing_2020},
particularly the Dark Energy SC (\desc), Transients and Variable Stars (TVS) SC, and Galaxies SC, but \pz metrics remain underdeveloped.
Some metrics are straightforward, such as the 10-year coadded depth or the number of supernovae with more than 10 epochs, and can be directly predicted from the \opsim metadata and visualized as sky maps with \healpix \citep{gorski_healpix_2005}.
However, \pz performance is quantified by derived statistics of a particular \pz estimator on an entire simulated galaxy catalog, often as a function of true redshift.
The holy grail of \os metrics would be a map of ``goodness of \pz quality" insensitive to \pz estimator for each \healpix pixel, a goal that has not yet been achieved, let alone with enough computational efficiency for practical \maf integration. 

Multiple aspects of the current approach to \pz metrics for \os optimization would benefit from improvement, ideally addressing as many as possible of the following needs:
\begin{itemize}
    \item An observing strategy metric for \pzs should be agnostic to the choice of estimation method as well as whether point estimates or \pzpdfs are used.
    \item A metric for \pz should be adaptable to integrate with metrics of 
    additional quantities sensitive to  observing strategy.
    \item An observing strategy metric for \pzs should not preclude direct comparison of overall metrics between science goals nor between analysis approaches for a shared science goal.
    \item Any metric included in the \maf should be fast and scalable; 
    simulation and propagation of mock data through an entire analysis pipeline is not feasible.
\end{itemize}
This paper explores a potential \os metric of estimated redshift quality that represents an improvement upon established metrics along the above axes.
Our metric relies directly on estimating the mutual information between photometry and redshift, in other words, quantifying how much information is gained on the redshift of galaxies by having access to the photometry under a given \os. 

This paper is not the first use of an information criterion for optimization in the context of redshift estimation.
\citet{malz_approximating_2018} used an information-theoretic metric to optimize the storage parameterization of \pzpdfs, however, \cite{kalmbach_info_2020} used a metric of mutual information more closely related to that of this work to optimize filter design for \pzs.
In their work, photometric data with Gaussian error was simulated using a simple redshift prior and a small number of galaxy SED templates.
This simple forward model provided an analytic method for calculating the mutual information.

In this paper, we leverage recent advances in machine learning to enable the calculation of the mutual information contained in more complex simulations for which an analytic model may be unavailable, as could be anticipated of even idealized data processed through \opsim under a realistically complex \os.
In Section~\ref{sec:method}, we introduce the mathematical framework for the variational mutual information lower bound that serves as the basis for our metric.
In Section~\ref{sec:demo}, we demonstrate the metric in the context of \os optimization for \lsst.
And in Section~\ref{sec:disco}, we summarize its strengths and future directions for its development and application.

\section{Method: Variational Mutual Information Lower Bound}
\label{sec:method}

This section introduces TheLastMetric (\tlmword), denoted as \tlmsymb,\footnote{the last letter of the Hebrew alphabet, pronounced ``tav"} an estimator-independent metric for parameters of interest conditioned on photometry under any \os.
Though we derive it in the context of redshift-dependent cosmological probes, its mathematical structure is not inherently restricted to redshift;
as it is so broadly applicable across science cases, one may jokingly exaggerate that it's the penultimate \os metric, hence its name.

In Section~\ref{sec:theory}, we introduce the mathematical formalism and review relevant information-theoretic concepts.
In Section~\ref{sec:metric} we derive the metric itself, and in Section~\ref{sec:nde} we describe the model by which the metric is calculated.

\subsection{An information theoretic view of observing strategies}
\label{sec:theory}

Every science case seeks a metric of how informative the data corresponding to each \os is with respect to some physical parameters of interest.
The properties of the telescope and its \os correspond to a transformation of the underlying true data in the universe.
Such true data would correspond to photons that could be observed only by an impossibly perfect, idealized instrument that collects complete, noiseless data, in contrast to what we can observe from any real telescope under a given \os, a subset of those photons restricted by the \os and convolved with instrumental errors.
Typically, the information content of recovered physical parameters is determined by running the observed data through an end-to-end analysis pipeline, which is impractical for a \maf.
This is particularly true for cosmological applications, but while our derived cosmological constraints depend on the analysis procedure \citep{chang_unified_2018}, the potentially recoverable cosmological information content does not.\footnote{Though we derive and demonstrate our metric with this application in mind, such a statement is no less true of other physical parameters of interest to a variety of science cases, making this metric broadly applicable beyond cosmology or even redshifts.}

Let us begin by quantifying the total cosmological information content of an idealized survey with perfect photometry using information theory. 
We denote the random variable representing cosmological parameters as $\Theta$ and a random variable representing survey photometry as $X_{phot}$. 
The information content about cosmological parameters $\theta \sim \Theta$ due to photometry $\data \sim X_{phot}$ can be described by the \textit{mutual information} between $\Theta$ and $X_{phot}$, defined as
\begin{equation}
I(\Theta \ ;  X_{phot})  \equiv \E_{p(\theta, \data)} \ \left[  \log\frac{p(\theta, \data)}{p(\theta) p(\data)}  \right] .
\label{eq:mutual_information}
\end{equation}
The mutual information between two random variables represents the reduction in uncertainty in one due to knowing the other. 
It is desirable for this mutual information to be large, i.e. we want the photometry $\data$ to be very informative about the cosmological parameters $\theta$.

In practice, of course, we do not have access to perfect photometry, and a typical cosmological analysis pipeline does not directly relate photometry to cosmology in a single step;
instead, typical analyses constitute a pipeline from observed photometry $X_{phot}^{obs}$ to cosmology $\Theta$ via a number of intermediate quantities. 
The relationship between cosmological parameters $\Theta$, redshift distribution $Z$, true photometry $X_{phot}^{true}$, and observed photometry $X_{phot}^{obs}$ is an example of a Markov chain $\Theta \rightarrow  Z \rightarrow X_{phot}^{true} \rightarrow X_{phot}^{obs}$ satisfying
\begin{equation}
    p(\theta, z, x_{phot}^{true}, x_{phot}^{obs}) = p(\theta) \  p(z|\theta) \  p(x_{phot}^{true} | z) \ p(x_{phot}^{obs} | x_{phot}^{true}) .
\end{equation}
The amount of information each stage of this chain retains about cosmology can be expressed using the \textit{data processing inequality} \citep{cover_elements_2006},
\begin{equation}
  I(\Theta \ ; \ Z) \geq I(\Theta \ ; \ X_{phot}^{true}) \geq I(\Theta \ ; \ X_{phot}^{obs}) \;,
\end{equation}
which can be intuitively understood as saying that information can only be lost through the steps of a chain. 

Every application of \lsst data aims to minimize the information loss in the steps of this chain, or one analogous to it, that are under our control.
An example of a step we as experimenters control is the choice of \os that min information loss in the stages of this chain that relates galaxies' true redshifts and their observed photometry $I(Z \ ; \  X_{phot}^{obs})$, which depends on the specific \os that transforms $X_{phot}^{true} \rightarrow X_{phot}^{obs}$. 
Again following the data processing inequality, we expect the following:
\begin{equation}
    I(Z \ ; \ X_{phot}^{true}) \geq I(Z \ ; \ X_{phot}^{obs}) \;,
\end{equation}
i.e. that the mutual information between redshift and observed photometry would be saturated if perfect, true, unobservable photometry were available, but the \os determines how closely we can approximate that bound.

Our goal is to compare the mutual information $I(Z \ ; \ X_{phot}^{obs})$\footnote{As was suggested in the preamble, this paper concerns the mutual information of redshift and photometry, but the derivation is just as valid for the mutual information $I(\Psi; X_{phot}^{obs})$ of observed photometry with some other parameter $\Psi$, for example e.g. stellar mass, used in any science application, even beyond cosmology.} for different \oss, with the understanding that the \os that can achieve the highest mutual information at this level of the chain will also maximize the overall mutual information with respect to cosmology $I(\Theta \ | \ X_{phot}^{obs})$ for a fixed analysis pipeline.
Given this scope, we henceforth use $X_{phot}$ as shorthand for $X_{phot}^{obs}$.
The next challenge to discuss is a practical computation of the mutual information.

\subsection{Tractable variational lower bound on the mutual information}
\label{sec:metric}

Evaluating the mutual information as defined in Equation~\ref{eq:mutual_information} is in general extremely challenging, as in most cases only samples from the distributions involved as accessible, not the underlying distributions themselves. 
In recent years however, the concept of mutual information has found many applications in the machine learning literature \citep[e.g.][as well as the recent review of \citet{Poole2019}]{Tishby2015, Alemi2016, Bachman2019}, which has driven significant research into tractable estimators of lower bounds.

To achieve a tractable expression for the mutual information $I(Z; X_{phot})$ of redshift and photometry, let us first introduce the \textit{entropy}
\begin{equation}
\label{eqn:entropy}
H(Z) \equiv -\int dz\ p(z) \log p(z)
\end{equation}
of the redshift distribution $p(z)$, which quantifies our uncertainty on the random variable $Z$.
Using Equation~\ref{eqn:entropy}, we rewrite the mutual information $I(Z; X_{phot})$ of redshift and photometry in the following way:
\begin{align}
\label{eqn:mutual}
I(Z ; X_{phot}) &= \E_{p(z, \data)} \ \left[  \log \frac{p(z | \data)}{p(z)} \right] \nonumber  \\
		 &= \E_{p(z, \data)} \left[ \log p(z | \data) \right]
		\nonumber  - \E_{p(z)} \left[ \log p(z) \right] \\
		 &= \E_{p(z, \data)} \left[ \log p(z | \data) \right] + H(Z) .
\end{align}
As we do not directly have access to the posterior distribution $p(z | \data)$, the first term is unknown, making the mutual information itself intractable.

However, the overall expression for the mutual information can be bounded from below by introducing a variational approximation $q_\varphi( z | \data)$ for the posterior $p(z | \data)$, which leads to the variational lower bound introduced in \cite{Barber2003}:
\begin{align}
\label{eq:bound}
	I(Z ; X_{phot}) &= \E_{p(z, \data)} \left[ \log \frac{p(z | \data)}{q_\varphi(z | \data )} \right]  \nonumber  \\ 
	 & + \E_{p(z, \data)} \left[ \log q_\varphi(z | \data)\right] + H(Z) \\
	    &= \mathcal{D}_{KL}\left[ p(z| \data)  || q_\varphi(z | \data)  \right] \nonumber\\
	 & + \E_{p(z, \data)} \left[ \log q_\varphi(z | \data ) \right] + H(Z) \label{eq:kl_term}
.
\end{align}
In this expression, $\mathcal{D}_{KL}[p(z | \data) || q_\varphi(z | \data)]$ is the Kullback-Leibler Divergence (KLD), a directional measure of the loss of information due to using the variational model $q_\varphi(z | \data)$ as an approximation to the true, unknown, posterior distribution $p(z | \data)$. 
Because this KLD is non-negative, this last expression can be used to provide the following lower bound on the mutual information:
\begin{align}
\label{eqn:tlm}
    I(Z; X_{phot}) &\geq \E_{p(z, \data)} \left[ \log q_\varphi(z | \data) \right] + H(Z) \equiv \tlmsymb,
\end{align}
thereby providing the definition of TheLastMetric (\tlmword).

Not only is this expression now tractable, as it can be estimated simply by \textit{optimization} of the variational parameters $\varphi$, but, moreover, the bound is tight when $q_\varphi(z | \data) = p(z | \data)$ is true.
Like the KLD, \tlmsymb\ has units of information, which are \textit{nats} in the base-$e$ convention used in this paper but can be trivially converted to, e.g. base-2 \textit{bits}.

\subsection{Lower bound implementation using conditional normalizing flows}
\label{sec:nde}

The lower bound on mutual information introduced in the previous section relies on having access to a parametric conditional density $q_\varphi(z | \data)$, also known as the variational distribution, which is optimized to match the true posterior $p(z | \data)$. 
Any parametric conditional density estimator could be used for this purpose, but the more expressive the model, the tighter the lower bound will be. 

In this work, we approximate $p(z | \data)$ with a Normalizing Flow (NF) \citep{rezende_nf_2015, dinh_nice_2015}, a flexible class of deep generative models that represents the current state-of-the-art on many density estimation tasks \citep{kobyzev_nfreview_2020}.
NFs are Latent Variable Models (LVMs), which model a given distribution $p(x)$ over a target variable $x$ by introducing 1) a latent variable $z$ that follows a known prior distribution (typically a multivariate normal distribution) $p(z)$ as $z \sim p(z)$ and 2) a parametric mapping $f_\varphi$ that maps this latent variable $z$ to a point in the target distribution $x$ according to $x = f_\varphi(z)$. 
This is no different from other deep generative latent variable models like Variational Autoencoders \citep{kingma_vae_2013} or Generative Adversarial Networks \citep{goodfellow_gan_2014}, but what sets NFs apart is that they are specifically designed using a \textit{bijective} mapping $f_\varphi$. 
In the case of a bijection, the probability density function $q_\varphi$ of the model can be expressed as
\begin{equation}
    q_\varphi(x) = p(z=f^{-1}_\varphi(x)) \left| \det \frac{\partial f_\varphi}{\partial x}(x) \right|^{-1}, \label{eq:nf_prob}
\end{equation}
where $\det \frac{\partial f_\varphi}{\partial x}$ is the Jacobian determinant of $f_\varphi$, which accounts for how $f_\varphi$ distorts volume elements. 
This expression is nothing more than the change of variable formula for probabilities, but it gives NFs a crucial advantage over other LVM models: their probability density function has an explicit closed-form expression. 
In other words, for a given set of parameters $\varphi$ we can explicitly compute the probability $q_\varphi(x)$ of a data point $x$ under the model. 

A Conditional NF (CNF) can be trivially made by introducing a conditional variable $y$ in the mapping $f_\varphi(z ; y)$ \citep{winkler_conditionalnf_2019} that preserves the bijectivity of the mapping for all $y$, so Equation~\ref{eq:nf_prob} still applies.
Thus the conditional distribution modeled by the flow is simply:
\begin{equation}
    q_\varphi(x | y) = p(z=f^{-1}_\varphi(x ; y)) \left| \det \frac{\partial f_\varphi}{\partial x}(x ; y) \right|^{-1}. \label{eq:nf_prob_cond}
\end{equation}
In practice, as the mapping $f_\varphi$ is typically implemented using a neural network, making a normalizing flow conditional simply amounts to adding the variable $y$ as an input to the networks parameterizing the flow.

Because (C)NFs have tractable likelihoods, they can be trained by directly optimizing the probability of the training set under the model. 
For CNFs, the training loss takes the following form
\begin{equation}
    \mathcal{L} = - \mathbb{E}_{p(x, y)} \left[ \log q_\varphi(x | y) \right],
    \label{eq:loss}
\end{equation}
which can be shown to minimize the KLD $\mathcal{D}_{KL}[p(x|y) || q_\varphi(x|y)]$, i.e. driving the (approximating) model distribution to be close to the (true) data distribution.
We note that the CNF's loss function given by Equation~\ref{eq:loss} is equal to the first term of the variational lower bound Equation~\ref{eq:bound}. 
In other terms, training the CNF with this loss function is exactly equivalent to maximizing the variational mutual information lower bound. 

The approach presented here to build a practical lower bound is agnostic to the NF architecture employed, but the specific choice should be appropriate to the details of the problem at hand. 
We defer the details of the model used in this work to Section~\ref{sec:code}.

\section{Demonstration in the context of redshifts for \lsst}
\label{sec:demo}

The necessary and sufficient conditions favoring the use of \tlmword are that it must be at least as effective as established metrics and have some additional advantage(s), such as utility, interpretability, and/or efficiency.
To demonstrate the former, we perform controlled experiments on appropriate data, described in Section~\ref{sec:data}, using an implementation of a variational approximation to $\tlmsymb$, presented in Section~\ref{sec:code}.
The latter is discussed in Section~\ref{sec:res}, where the restults of this experiment are presented.

\begin{figure}
  \centering
  \includegraphics[width=0.45\textwidth]{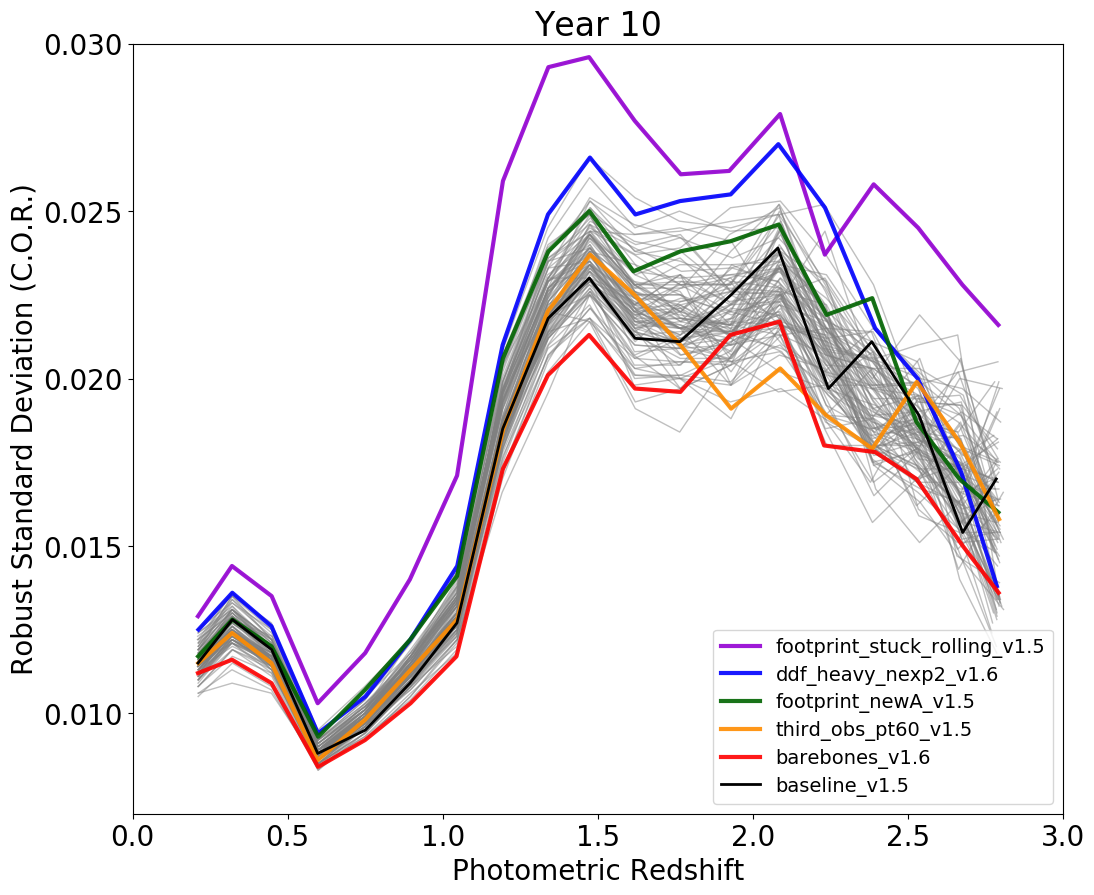}\\
  \includegraphics[width=0.45\textwidth]{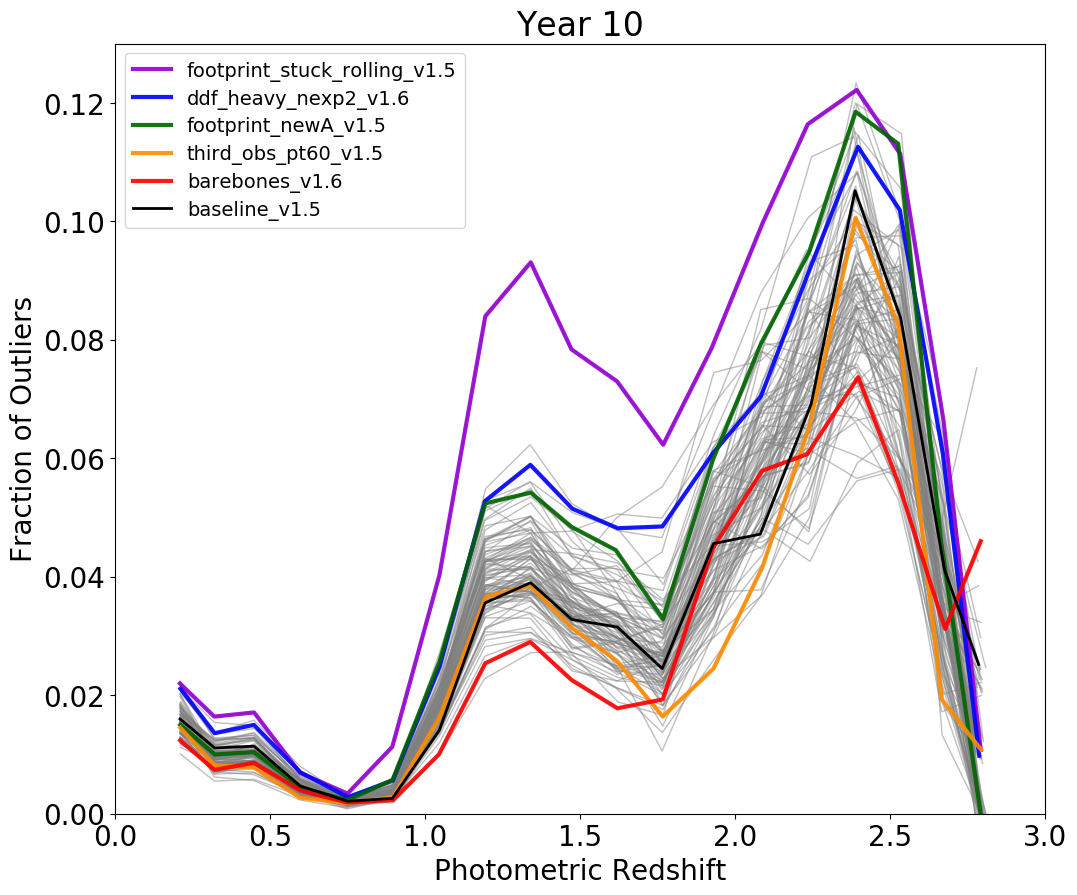}
  \caption{\cmnn-estimated \pz statistics for a wide variety of \opsim conditions (grey), the baseline \opsim (black), and the exemplary \opsim \os simulations used in this pilot study (colors as in legend).
  The \base \os has mid-quality results, and the simulations used in this analysis were chosen because they represent a range of \pz qualities that stand out as better and worse than \base.
  These results should not be taken as representative of the future \textit{absolute} quality of \lsst \pz, but the differences in the results are representative of the impact of different \opsim strategies on the \pz results.}
  \label{fig:photoz} 
\end{figure}

\begin{table*}
\centering
\begin{tabular}{l|l|l|l|l|l|l}
\opsim \os Simulation &  m5\_u  &  m5\_g  &  m5\_r  &  m5\_i  &  m5\_z  &  m5\_y\\
\hline
baseline\_v1.5                   & 25.86 & 27.02 & 26.99 & 26.42 & 25.70 & 24.94 \\
footprint\_stuck\_rolling\_v1.5  & 25.56 & 26.68 & 26.62 & 26.06 & 25.33 & 24.61 \\
ddf\_heavy\_nexp2\_v1.6          & 25.57 & 26.82 & 26.84 & 26.26 & 25.57 & 24.82 \\
footprint\_newA\_v1.5            & 25.75 & 26.87 & 26.85 & 26.29 & 25.55 & 24.78 \\
third\_obs\_pt60\_v1.5           & 25.87 & 27.03 & 26.99 & 26.43 & 25.70 & 24.93 \\
barebones\_v1.6                  & 26.00 & 27.13 & 27.07 & 26.57 & 25.78 & 25.05 \\
\end{tabular}
\caption{
	Simulated median 5$\sigma$ limiting magnitudes in extragalactic regions for coadded images from the wide-fast-deep survey, from the \base \opsim \os simulation and the five exemplary \opsim \os simulations used in this work.
}
\label{tab:limmag}
\end{table*}

\begin{figure}
\centering
\includegraphics[width=0.45\textwidth]{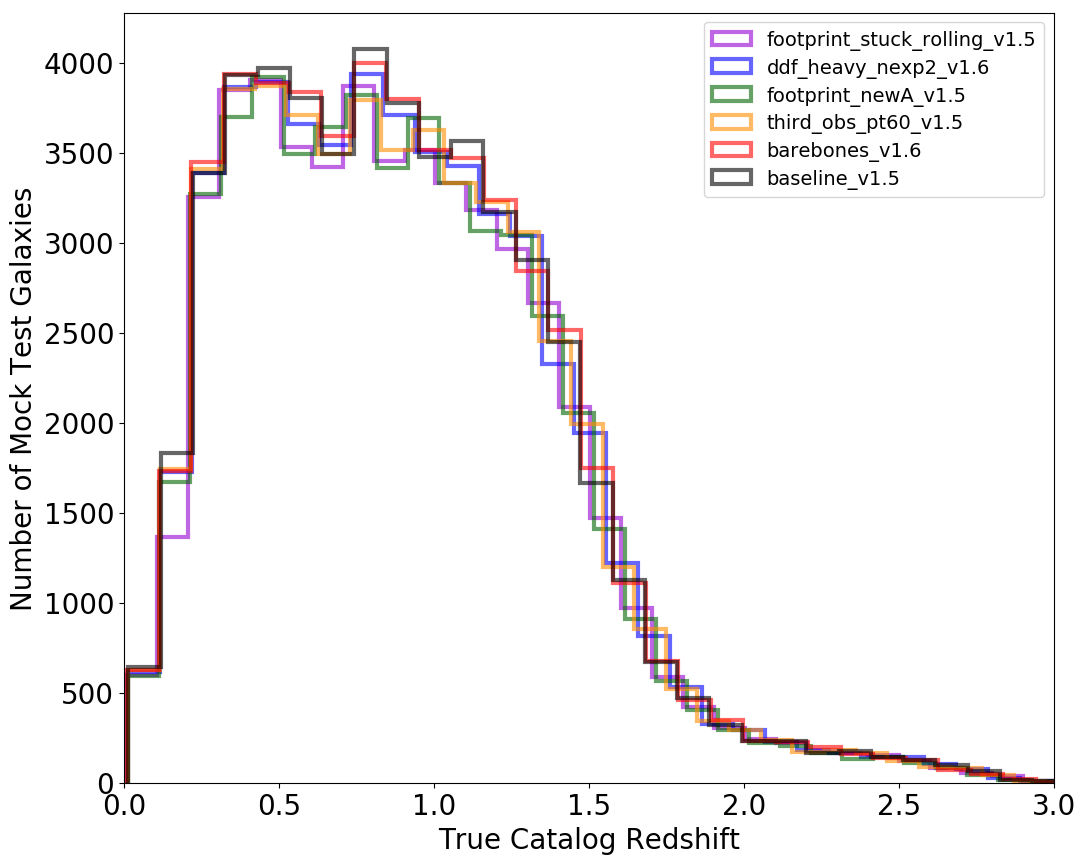}
\caption{
	The true catalog redshift distribution for the mock test galaxy sets used to simulate \pz results for each \opsim \os simulation.
	Differences between the redshift distributions across \opsim \os simulations are due to statistical fluctuations and correspond to a $0.5\%$ difference in entropy $H(Z)$, meaning they are statistically indistinguishable.
}
\label{fig:redshift_hist} 
\end{figure}

\subsection{Data: simulated \pz catalogs}
\label{sec:data}

For this work, we use simulations based on the same mock galaxy catalog to which the Color-Matched Nearest-Neighbors (\cmnn) \pz estimator\footnote{A demonstration of the \cmnn \pz estimator is available on {\tt GitHub} at \url{https://github.com/dirac-institute/CMNN_Photoz_Estimator}.} was applied to produce \lsst-like \pz results in \citet{graham_photometric_2018, graham_photometric_2020} and the same set of \opsim\ conditions \citet{lochner_impact_2021} used for \desc's assessment of the impact of \os on multi-probe cosmological constraints.
As an overview, we use \cmnn to generate mock photometry for a given \opsim simulation, then use the traditional \pz metrics evaluated on \cmnn's \pz estimates to identify exemplary \oss to which we then apply \tlmword.

For each \opsim\ \os simulation, we first determine the 5$\sigma$ limiting magnitude of the 10-year coadded images from the wide-fast-deep program in sky regions (\healpix $\sim$220 arcmin wide).
We identify regions as extragalactic fields (i.e., appropriate for cosmological studies) if their Galactic dust extinction was E(B-V)$<$0.2 mag and if they received at least 5 visits per year in all six $ugrizy$ filters, and then calculate the median 10-year depths over all extragalactic fields, which are reported in Table~\ref{tab:limmag}.
For every \opsim\ \os simulation, these 5$\sigma$ depths in each filter, $m_5$, are passed to the \cmnn Estimator, which uses them to calculate magnitude errors for a catalog of mock galaxies: 
$\sigma_{\rm rand}^2 = (0.04-\gamma)x + \gamma x^2$, where $x=^{0.4(m_{\rm true}-m_5)}$, $m_{\rm true}$ is the true catalog apparent magnitude, and $\gamma$ is a filter-dependent factor which account for the effect of, e.g., sky brightness \citep[see Section 3.2 of ][]{2019ApJ...873..111I}.
The \cmnn\ Estimator then simulates observed apparent magnitudes by drawing a random value from a normal distribution with a standard deviation equal to $\sigma_{\rm rand}$ and adding it to the true catalog magnitude.
Test- and training-sets are drawn randomly (without replacement) from the mock galaxy catalog, and \pz\ estimates for the test set are generated by identifying training-set galaxies with similar colors, i.e. a subset of color-matched nearest-neighbors in 5-dimensional color space, and adopting as the true redshift one test-set galaxy's \pz chosen at random. 
In order to standardize the mock catalogs used for all simulations, we applied the same cuts on the observed apparent magnitudes of 25.0, 26.0, 26.0, 25.0, 24.8, and 24.0 mag in filters {\it ugrizy} to both the test and training sets, about half a magnitude brighter than the brightest 5$\sigma$ limiting depth of any given \opsim\ \os simulation, as in \citet{lochner_impact_2021}.

It is important to note that for all of our simulations, the test- and training-sets were drawn from the same intrinsic mock catalog, which means they are perfectly matched in terms of their redshift and apparent magnitude distributions.
While this would be concerning if we aimed to evaluate the realistic performance of the \cmnn estimator, its role in this study is not to produce \lsst's ``official" or even ``best" \pzs;
rather, we use it as a forward model of the relationship between the relative quality of \pz estimates given the quality of input photometry, which provides us with mock photometry under a given \os upon which we demonstrate \tlmword.

Several statistical measures are commonly used to evaluate the quality of test-set \pz estimates based on
the \pz\ error, $\Delta z \equiv (z_{true}-z_{phot}) / (1+z_{phot})$.
While the bias $\langle \Delta z \rangle$, representing systematic over- or under-estimates of point estimates, was calculated, the values under these idealized conditions of a perfectly representative and complete training-set were too small and similar for bias to discriminate between \oss.
However, our experimental design does produce simulated \pz\ results for which the standard deviation and fraction of outliers are relatively improved or degraded in a way that correlates with the 5$\sigma$ depths in the six \lsst filters.
For a robust standard deviation in $\Delta z$ we use the interquartile range (IQR) divided by 1.349, making the fraction of outliers the fraction of test-set galaxies with $|\Delta z|>3\times$ the robust standard deviation or $>3\times0.06$, whichever is larger \citep[matching the definition of the \lsst Science Requirements Document,][]{LSST2013}.

In Figure~\ref{fig:photoz}, we show the robust standard deviation and fraction of outliers\footnote{See \cite{graham_photometric_2020} for a full description of the standard deviation and fraction of outliers statistics.} in bins of \pz\ for a wide variety of \opsim\ \os simulations, highlighting the baseline \opsim\ \os simulation and a selection of five additional \opsim\ \os simulations which produced notably better or worse results than the baseline that we use in this work, whose median 10-year depths are provided in Table~\ref{tab:limmag}.

In Figure~\ref{fig:redshift_hist}, we show the distribution of true catalog redshift for the test-set galaxies used for each simulation.
Because the same magnitude cuts were applied to the test and training sets for all simulations, the training sets have similar redshift distributions.
The differences between the lines are just the statistical random fluctuations caused by drawing a new test subset (of 50000) from the greater mock galaxy catalog (of millions of galaxies) for each simulation.
The drop in the number of galaxies in high redshift bins is realistic for the applied cuts on apparent magnitude, and is the cause for some of the observed scatter in the \pz\ statistical results seen in Figure~\ref{fig:photoz}.

\begin{figure}
    \centering
    \includegraphics[width=0.35\textwidth]{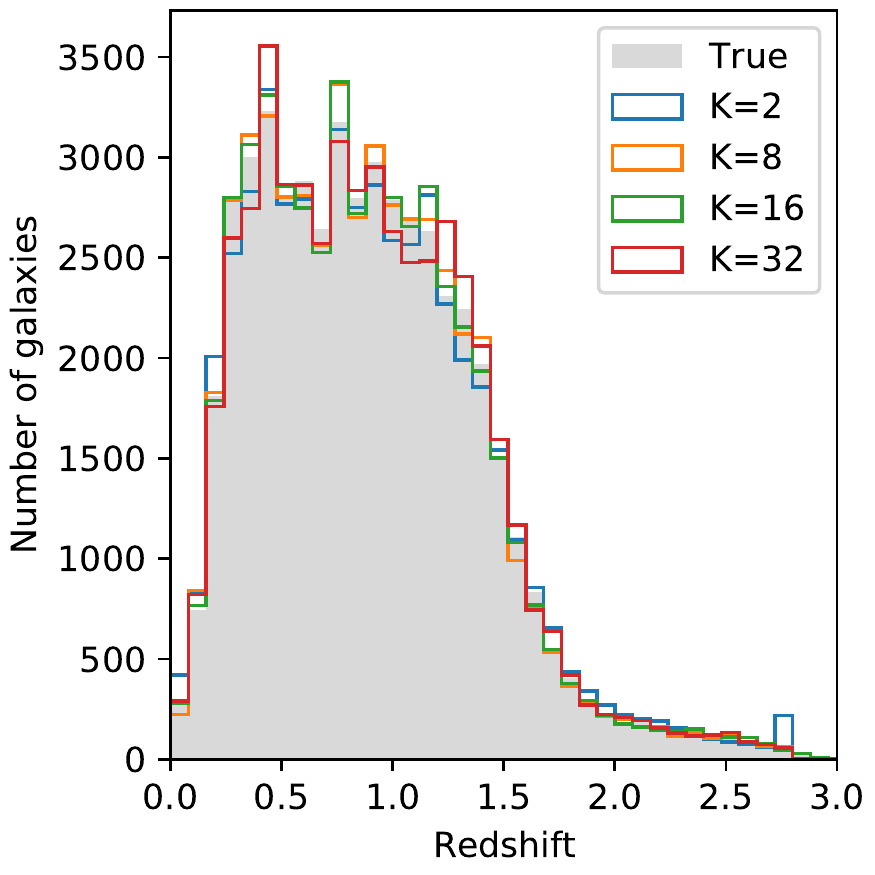}
    \caption{Histogram of the redshift distributions for \base under normalizing flows with different values of the tuning parameter $K$.
    The true distribution is in gray, while the distribution learned by normalizing flows with various values of $K$ are displayed in color.
    Aside from a high-redshift artifact for the extreme $K=2$ flow, the choice of $K$ appears to have little effect.}
    \label{fig:K-dep}
\end{figure}

\subsection{Implementation: density estimation with \pzflow}
\label{sec:code}

We build the NF lower bound discussed in Section \ref{sec:nde} using \pzflow \citep{crenshaw_pzflow_zenodo_2021}, a GPU-enabled python package for normalizing flows built on \jax \citep{jax_github_2018}.
We refer to our public implementation as \tlmcode\footnote{\url{https://github.com/aimalz/TheLastMetric}}.

For the latent distribution $p(z)$, we use the uniform distribution $\mathcal{U}(0, 3.2)$, as sampling and density estimation are trivial, and the domain matches the compact support of the redshifts in our data set.
Matching the features of the latent space to the data eases training and prevents any potential unphysical outliers.

For the bijection $f_\varphi$, we use a rational-quadratic neural spline coupling \citep[RQ-NSC;][]{durkan_rqnsf_2019}, which is a state-of-the-art bijection both capable of modeling high-dimensional distributions with hundreds of modes and efficient at both sampling and density estimation.
The RQ-NSC transforms galaxy redshifts with monotonically-increasing piecewise combinations of $K$ segments, each of which is a rational-quadratic function.
The bijection parameters $\varphi$ are the values and derivatives of these rational-quadratic functions at $K+1$ spline knots.
The value of $K$ impacts the resolution of the distribution learned by the NF, with large $K$ corresponding to high resolution.
After fixing $K$, a neural network calculates the values and derivatives of the $K+1$ knots from the conditional variables: 
the five \lsst colors ($u-g$, $g-r$, $r-i$, $i-z$, $z-y$) and the $r$ band magnitude (which serves as a proxy for overall luminosity).
After assessing several configurations for the neural architecture of our NFs, we adopted a single RQ-NSC coupling layer, parameterized by a dense neural network with 2 layers of size 128 and ReLu non-linearities. 

To confirm robustness to the choice of tuning parameter $K$, we train flows with $K=$ 2, 8, 16, and 32 for each of the six \oss under consideration.
Each flow is trained by minimizing the loss of Equation~\ref{eq:loss} with respect to the parameters $\varphi$ using the \code{Adam} optimizer \citep{kingma_adam_2014}.
We train with a learning rate (lr) of $10^{-3}$ for 100 epochs, followed by lr $=2 \times 10^{-4}$ for 100 epochs, followed by lr $=10^{-4}$ for 50 epochs.
For each flow, this takes about 1 minute on a Tesla P100 12GB GPU (or about twice that on a CPU).
Figure~\ref{fig:K-dep} shows the redshift distributions learned by the four flows trained on the \base \os.
Aside from a high-redshift artifact for $K=2$, the choice of $K$ has little effect on the redshift distribution learned by the flow.
Similar behavior is observed for the redshift distributions of the other \oss, as well as the cross-correlations with galaxy colors.
Thus, for the remainder of this work, we will just use $K=16$.

Now that we have set the value of $K$, for each \os we train nine additional flows with $K=16$ and the same training schedule.
The 10 flows per \os will be used as a deep ensemble \citep{deep_ensembles_2016} to account for the \textit{epistemic uncertainty} of the model.
Deep ensembles perform approximate Bayesian marginalization \citep{wilson_bayesian_2020} over network parameters by independently initializing and training neural network parameters a number of different times.
In the case of a non-convex loss function, this procedure often results in solutions that live in distinct basins of attraction in parameter space \citep{fort_deepEnsembles_2020} and is therefore preferable to methods that approximately marginalize over single basins of attraction, such as the Laplace \citep{mackay_laplace_1992} and SWAG \citep{maddox_swag_2019} approximations.
We calculate a distribution and report a mean of \tlmsymb for each \os based on the ten trained NFs.

\begin{figure}
  \centering
  \includegraphics[width=0.45\textwidth]{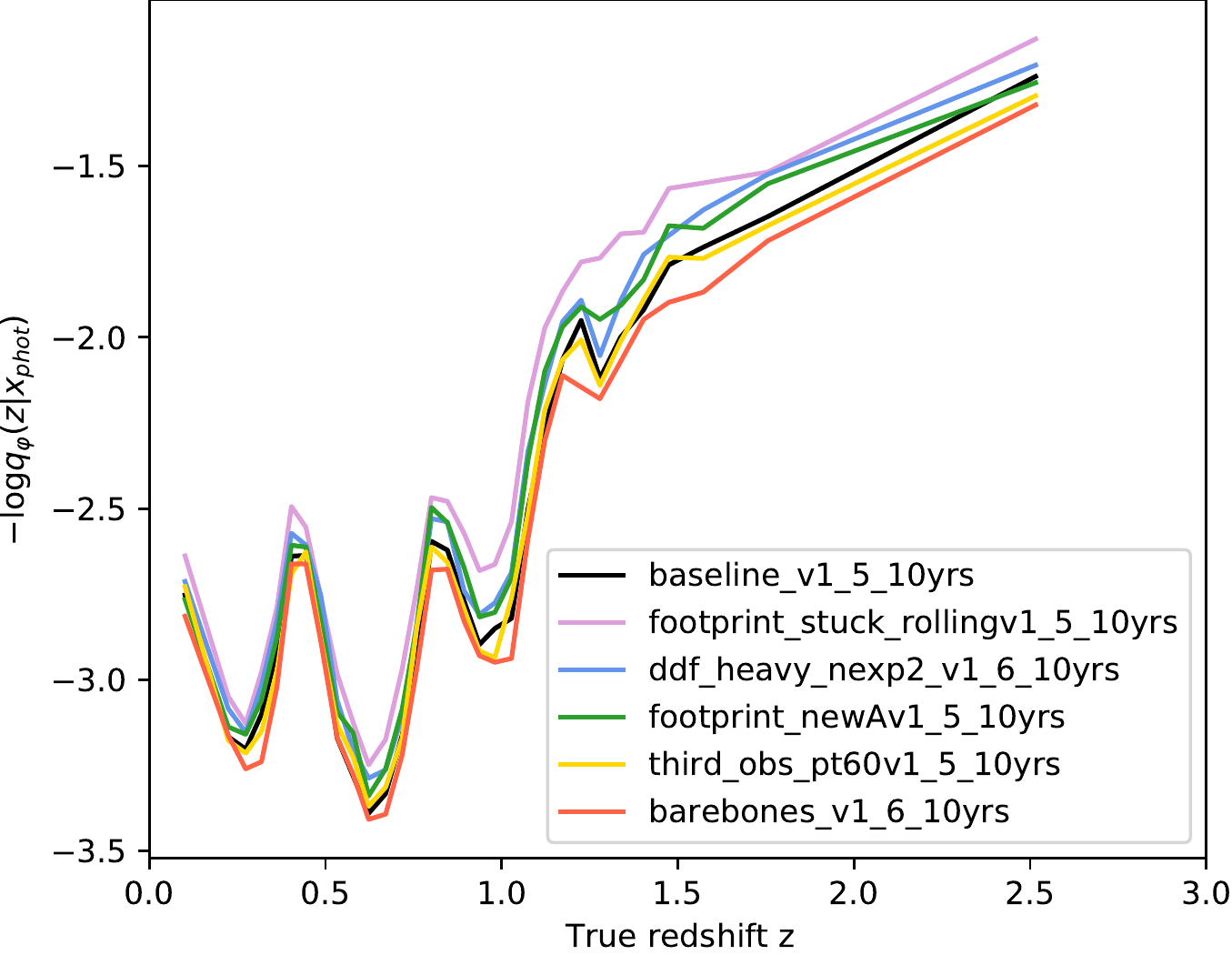}
  \caption{
  Redshift-binned average of negative log-posterior $- \log q_\varphi (z | x_{phot})$ 
  under each \opsim\ \os simulation, analogous to Figure~\ref{fig:photoz} (i.e. lower is better). 
  This component of \tlmword indicates the uncertainty on redshifts given observed photometry, for instance, the reduction in redshift uncertainty as the Balmer break passes between the \lsst photometric filters, and otherwise follows the dominant trend of the traditional \pz metrics by worsening at high redshift.
  Though the general ranking of \oss matches that of the traditional metrics, the reorderings between redshift bins is not as severe, an indication of robustness of \tlmword.
}
  \label{fig:metric_hist_redshift} 
\end{figure}

\begin{figure}
    \centering
    \includegraphics[width=0.45\textwidth]{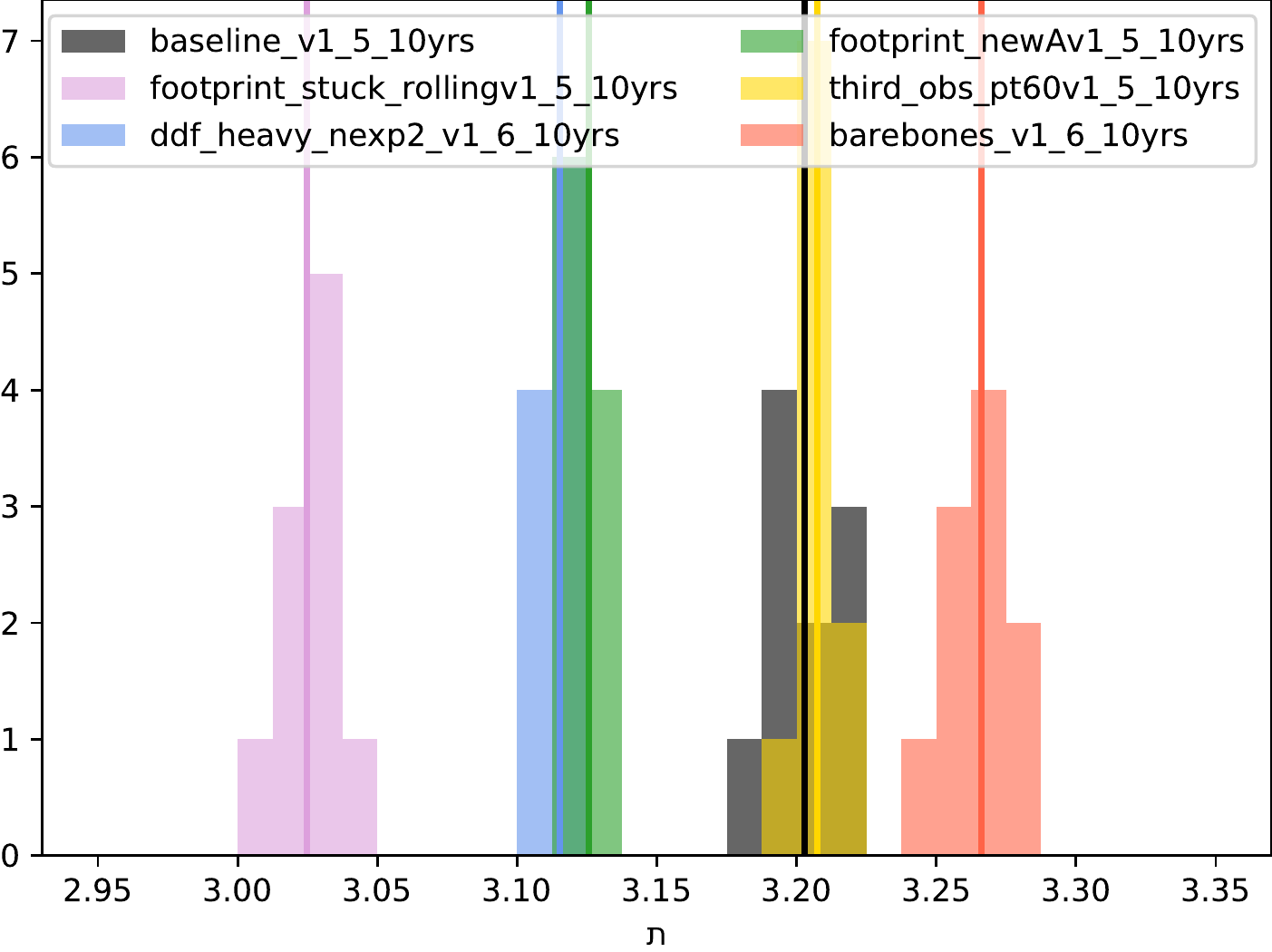}
    \caption{The distribution of \tlmsymb estimates for each \os. 
    The shaded distributions represent an estimate of the epistemic errors in the evaluation of the metric, obtained using a deep ensemble approach. 
    Mean \tlmsymb estimates indicated by vertical lines are obtained by averaging of the deep ensemble values.
    The stochasticity exceeds the difference between metric values for two pairs of \oss, but stratification by \tlmword is otherwise robust.
}
    \label{fig:metrics_consistency}
\end{figure}

\begin{figure}
  \centering
  \includegraphics[width=0.4\textwidth]{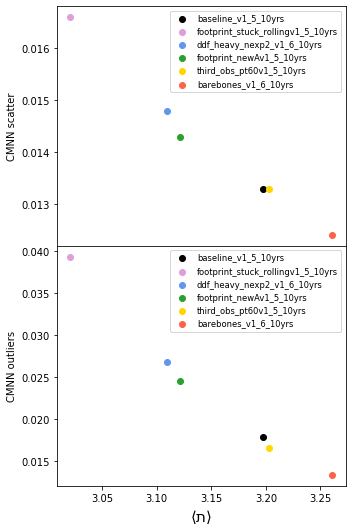}
  \caption{
  A comparison between the mean \tlmsymb\ and the traditional \pz metrics of intrinsic scatter (top panel) and outlier rate (bottom panel) from the \cmnn estimator, calculated for the redshift range $0.3 < z_{\rm phot}<3.0$ for each \os (colors).
  Both \tlmword and traditional \pz metrics penalize the \fpstuck and \ddf \oss and favor \bare and generally agree on the ranking of \oss. 
}
  \label{fig:metric_comp} 
\end{figure}

\subsection{Results}
\label{sec:res}

As a hypothesis, we expect that \tlmword will confirm the hierarchy of \oss corresponding to the traditional \pz\ statistics;
as will be discussed in Section~\ref{sec:disco}, the advantages of \tlmword are its potential extensions, but we first establish its consistency with our intuition based on the established \pz metrics.

Before considering \tlmword's value for each \os, we aim to build some intuition of how it behaves in practice. 
We therefore begin by considering the behavior of a key component of \tlmsymb: 
the per-galaxy log-posterior $\log q_\varphi(z | x_{phot})$, which quantifies how probable the true redshift of a galaxy with photometry $\data$ is under the approximated posterior redshift distribution. 
This value would be high for narrow posteriors centered on the true redshifts, indicating that the photometry is very constraining of redshift.
Alternatively, a low value indicates that the posterior is not very concentrated and/or offset from the true redshift, meaning that the photometry is not very constraining.

Figure~\ref{fig:metric_hist_redshift} shows the redshift-binned negative expected value $\langle - \log q_\varphi(z | x_{phot}) \rangle$ for different \oss;
the minus sign is included for easier comparison with Figure~\ref{fig:photoz}, i.e. lower is better. 
We confirm the conclusions of the \cmnn statistics of Figure~\ref{fig:photoz}: 
towards higher redshifts, photometry becomes less constraining, in the same way that the scatter and outlier rate increases for \cmnn, and the \bare \os consistently outperforms the \fpstuck \os at all redshifts.
We also observe from Figure \ref{fig:metric_hist_redshift} that the ordering of \oss by these curves does not significantly depend on redshift, which is largely consistent with the findings of the traditional metrics of Figure~\ref{fig:photoz}.
This indicates that there is not a sub-range of redshifts for which a given strategy would outperform the others, which implies that a single \os could be optimal for both low-redshift and high-redshift science use cases.

In addition, $\langle - \log q_\varphi(z | x_{phot}) \rangle$ achieves a series of local minima, corresponding to increased information content, around the redshifts where the 4000 \AA\ Balmer break, a broad feature of galaxy spectra important for \pz estimation, crosses between the \lsst photometric filters.
This provides a strong consistency check that our implementation of the metric is capturing the real, physical mutual information between photometry and redshift.
It also demonstrates consistency with the results of \cite{kalmbach_info_2020}, who found that using information theory to optimize filters for \pzs corresponds to designing filters that can optimally constrain the location of the Balmer break as it moves across the optical wavelength range.

As described in Section~\ref{sec:metric}, \tlmsymb itself is obtained by taking the expectation of the log-posterior $\log q_\varphi(z | x_{phot})$ over the entire sample, thus capturing how informative the observed photometry is about the redshifts across the whole population, and adding an entropy term $H(Z)$ which only depends on the redshift distribution.

The distribution of \tlmsymb from the deep ensembles for each \os are shown in Figure~\ref{fig:metrics_consistency}, recalling that higher \tlmsymb is better.
The epistemic uncertainty in \tlmcode dominates \tlmsymb's discriminatory power for two pairs of \oss, but there is a clear four-tiered hierarchy of the redshift information each \os's photometry preserves.

Figure~\ref{fig:metric_comp} shows the mean values $\langle \tlmsymb \rangle$ of these distributions, plotted against the two relevant canonical \pz metrics of Figure~\ref{fig:photoz} for each \os, confirming that the behavior of \tlmsymb is qualitatively similar to that of the traditional \pz metrics\footnote{It also confirms the close correlation between the traditional intrinsic scatter and outlier rate, as the latter is defined in terms of the former.}, favoring \bare and disfavoring \fpstuck.

\section{Discussion \& conclusions}
\label{sec:disco}

In this paper, we introduce \textit{TheLastMetric}, \tlmsymb, a metric of mutual information, and apply it in the context of observing strategy optimization for an astronomical survey with diverse scientific goals.
We also present \tlmcode to the community, an implementation of a variational lower bound on the mutual information of photometry with respect to redshift.
We demonstrate the calculation of \textit{TheLastMetric} on mock photometric galaxy catalogs corresponding to exemplary observing strategies for \lsst, confirming that it is qualitatively similar to conventional \pz metrics.

\textit{TheLastMetric} offers distinct advantages addressing key needs for observing strategy metrics for \lsst's diverse extragalactic goals:
\begin{itemize}

    \item \textit{TheLastMetric} is a measure of information in units of nats, meaning it and any science-case specific extensions thereof are directly comparable, enabling the isolation of the relative importance of science goals from the raw values of their observing strategy metrics.
    \item \tlmcode does not assume any \pz estimator, freeing it from assumptions of \pz template libraries, training sets, and other priors, as well as from the computational overhead associated with many popular \pz estimators.
    \item \textit{TheLastMetric} is applicable across redshift-dependent science cases as well as to other quantities informed by photometry that is influenced by observing strategy.
\end{itemize}

\textit{TheLastMetric} is not without its own assumptions of course.
We show that it is robust to the tuning parameters of the \tlmcode back-end, but evaluation on draws from the conditional normalizing flow model rather than the same data upon which it was trained would, strictly speaking, be more self-consistent.
Furthermore, though \tlmcode eliminates the computational expense of estimating \pzs, it retains the traditional metrics' computational overhead of simulating a mock galaxy catalog from \opsim parameters.

We note that the entropy $H(Z)$ of the redshift distribution of the mock galaxy catalog, which factors into \tlmsymb in Equation~\ref{eqn:tlm} and thus influences Figures~\ref{fig:metrics_consistency} and ~\ref{fig:metric_comp}, may differ between cosmological probes or other science cases that use subsamples of galaxies under different selection functions.
Though the entropy term $H(Z)$ would need to be recomputed for the anticipated redshift distribution of the science-motivated subsample, that term is subdominant in magnitude as well as trivial to calculate.
Since the expected value of Equation~\ref{eqn:tlm} is defined in terms of $p(z, \data)$, \tlmsymb could be recalculated under a different redshift distribution without requiring retraining of \tlmcode to obtain the $\log q_\varphi(z | x_{phot})$ for each mock galaxy.
Thus \textit{TheLastMetric} is extensible to redshift-dependent science cases without increasing computational expense beyond what is required of the current \pz metrics.

Mathematically, \textit{TheLastMetric} is not inherently exclusive to redshift and may be extended to any parameter of interest available in the truth catalog of a mock galaxy sample to yield an interpretable metric of how informative the photometry is about that parameter.
Use of \textit{TheLastMetric} and potential extensions thereof, within and beyond cosmological applications, will enable the identification of an appropriate observing strategy for \lsst. 
We thus recommend \tlmcode's inclusion in the \maf and motivate future development into further mutual information metrics specific to individual science cases or probes of a single science application.

\section*{Acknowledgements}

This work was incubated at the August 2020 TVS SC \maf Hackathon\footnote{\url{https://lsst-tvssc.github.io/metricshackathon2020}}, which was supported by an LSSTC Enabling Science small programs grant.

AIM acknowledges support from the Max Planck Society and the Alexander von Humboldt Foundation in the framework of the Max Planck-Humboldt Research Award endowed by the Federal Ministry of Education and Research.
JFC is supported by the U.S. Department of Energy, Office of Science, under Award DE-SC-0011635, as well as the National Science Foundation, Division Of Astronomical Sciences, under Award AST-1715122, and the Office of Advanced Cyberinfrastructure, under Award OAC-1739419.
MLG acknowledges support from the DIRAC Institute in the Department of Astronomy at the University of Washington.
The DIRAC Institute is supported through generous gifts from the Charles and Lisa Simonyi Fund for Arts and Sciences, and the Washington Research Foundation.

\bibliographystyle{mnras}
\bibliography{ms}

\bsp
\label{lastpage}
\end{document}